\documentclass{llncs}

\usepackage{amsmath,amsfonts,datetime,listings,xspace}
\usepackage{color}
\usepackage{url}

\usepackage[noend]{algorithm2e}
\SetAlgoSkip{}

\newtheorem{thm}{Theorem}
\newtheorem{lem}{Lemma}

\newcommand{\Pex}{\textsc{Pex}\xspace}
\newcommand{\Klee}{\textsc{KLEE}\xspace}
\newcommand{\CBATool}{\textsc{CBA}\xspace}

\newcommand{\tout}{\textbf{timeout}}

\newcommand{\chn}[1]{\ensuremath{\mathbf{#1}} }

\renewcommand{\phi}{\varphi}
\newcommand{\N}{\mathbb{N}}
\newcommand{\cS}{\mathcal{S}}
\newenvironment{mitemize}{\vspace{-1\baselineskip}\begin{itemize}
\setlength{\itemsep}{0pt}}{\end{itemize}}

\renewcommand{\paragraph}[1]{\vspace{0.2cm}\noindent{\bf #1}~~}

\usepackage{times}
\usepackage[margin=1.52in]{geometry}

\begin{document}

\title{Efficient Loop Navigation for Symbolic Execution}

\author{Jan Obdr\v{z}\'alek and Marek Trt\'ik}

\institute{Masaryk University, Brno, Czech Republic
 \\\email{\{obdrzalek,trtik\}@fi.muni.cz} 
}

\maketitle

\begin{abstract}
  Symbolic execution is a successful and very popular technique used in
  software verification and testing. A key limitation of symbolic execution
  is in dealing with code containing loops. The problem is that even a
  single loop can generate a huge number of different symbolic execution
  paths, corresponding to different number of loop iterations and taking
  various paths through the loop.

We introduce a technique which, given
a start location above some loops and a target location anywhere below these
loops, returns a feasible path between these two locations, if such a path exists.  The
technique infers a collection of constraint systems from the program and
uses them to steer the symbolic execution towards the target. On reaching a
loop it iteratively solves the appropriate constraint system to find out
which path through this loop to take, or, alternatively, whether to continue
below the loop. To construct the constraint systems we express the values of
variables modified in a loop as functions of the number of times a given
path through the loop was executed.

We have built a prototype implementation of our technique and compared it to
state-of-the-art symbolic execution tools on simple programs with loops. The results show
significant improvements in the running time. We found
instances where our algorithm finished in seconds, whereas the other
tools did not finish within an hour. Our approach also
shows very good results in the case when the target location is not
reachable by any feasible path.
\end{abstract}

\section{Introduction}
Symbolic execution has been studied since 70's \cite{BEL75,Kin76}. The
main idea of symbolic execution is to represent input by symbols and then to
symbolically perform operations on values dependent on these
inputs. I.e. symbolic execution is a generalization of concrete
execution. Symbolic execution is usually used in the context of automatic
test generation.  With the arrival of powerful SMT
(Satisfiability-Modulo-Theories) solvers, e.g.~\cite{Z3,STP},
came a generation of powerful software tools for verification and test
generation \cite{TdH08,CDE08,NRTT09,SMA05}.

However symbolic execution quickly reaches its limits when confronted with
loops. As loops are widely used this is a significant problem. A typical
situation is that reaching a particular location below a loop depends on the
number of times this loop was iterated. Even worse, reaching that location
may depend not only on the number of iterations, but also on what particular
paths through the loop were chosen, and the order in which they were
taken. Since in symbolic execution any iteration of a loop creates a new
branch in the tree of symbolic executions, the size of the tree can become
very large with even a single loop. Without deriving any information about
the loop symbolic execution is forced to systematically explore all branches
of this tree, running out of time even on small programs.

We aim to solve the following problem: Given
a start location above a piece of code containing complicated loops,
including loop sequences and loop nesting, and a target location anywhere in
the code below, the goal is to find some feasible path between the start and
target location, if such a path exists.

The idea behind our algorithm is relatively simple. Assume we perform symbolic execution and want
to reach a given target location. On reaching a loop we enquire an oracle
which paths through this loop, and in which order, we should execute in
order to reach the target location. Following the oracle's advice we get to
our target, building path condition along the way. Only in our approach
the oracle is replaced by a constraint system, which is less powerful. For
each iteration it may suggest the next path to take, or to finish iterating
this loop.

To build the constraint system we express the values of variables modified
in a loop as functions of the number of times a given path through the loop
was executed. This concept extends the simple one of counting loop
iterations. Moreover, multiple counters for each path through the loop may be
needed to correctly handle loop nesting. The expressed values are then 'merged' over
all paths through a given loop. Constraint system is then created by taking 
branching conditions of the code below the loop and replacing all variables
with the corresponding functions of loop counters.

We suggest that our algorithm is most useful when integrated into existing
tools based on symbolic execution. Since the algorithm is itself based on
symbolic execution the integration should not be a big problem. Our
algorithm would work as a specific search strategy, activated when a global
search strategy needs to navigate to a specific program location below some
complicated loop structure. This would greatly improve the loop handling
ability of such symbolic execution tool and allow it to explore more code in
less time.

To evaluate our approach, we have built an experimental implementation of
our technique -- a tool called \CBATool.
We tested \CBATool on nine benchmarks we designed to capture those loop
structures which often appear in practice. 
\CBATool was able to solve all the benchmarks in seconds, confirming that the
approach we decided to use is correct. We also compare the performance of
\CBATool to successful symbolic execution tools
\Pex~\cite{TdH08,Pex} and \Klee~\cite{CDE08} and show that, on our set of
benchmarks, \CBATool is several orders of magnitude faster than either of
these tools.

The rest of the paper is organized as follows. In Section~\ref{sec:Overview}
we recall the basics of symbolic execution and then show how our technique
works on an example. The algorithm is explained in 
Section~\ref{sec:Algorithm}. 
In Section~\ref{sec:Results} we first describe the set of benchmarks used for testing
loop handling capabilities and compare the running times of \CBATool to
those of other tools. We also evaluate the performance data from various stages
of our algorithm. We survey related work in Section~\ref{sec:Related}, and
conclude with Section~\ref{sec:Conclusions}.


\section{Overview}
\label{sec:Overview}
In this section we start with an example showing the limitations of symbolic
execution when it comes to dealing with loops. We then present an overview
of our approach to solving this problem on this example. Detailed
description of our algorithm is deferred to the next section.

\subsection{Symbolic Execution and Loops}
\label{sec:Overview_1}

The idea behind symbolic execution can be explained as follows: Instead of
executing the program on a concrete input, we introduce for each input variable
\texttt{i} a symbolic value $\alpha_\mathtt{i}$, standing for some
concrete, but yet unknown value from the domain of \texttt{i}. The execution
of the program then proceeds in much the same way as normal (concrete) execution. The
important difference is that now many variables can  contain symbolic
values. For example let \texttt{a} and \texttt{b} be input variables with
symbolic values $\alpha_\mathtt{a}$ and $\alpha_\mathtt{b}$. Then  after
executing the statement \texttt{c = 2*a + b} the variable \texttt{c} will
contain the symbolic  value $2\cdot\alpha_\mathtt{a}+\alpha_\mathtt{b}$.

Branching is treated in the following way. During symbolic execution we
maintain a boolean formula $\phi$ called the \emph{path condition},
originally set to \texttt{true}. Assume that the symbolic execution reached a
branching statement and let $\psi$ be the associated condition. If
$\phi\wedge\psi$ is satisfiable we continue with the true branch and we
update the path condition to $\phi\wedge\psi$. (Similarly for the condition
$\phi\wedge\neg\psi$ and false branch.) If both $\phi\wedge\psi$ and
$\phi\wedge\neg\psi$ are satisfiable we fork the execution and work on the
resulting branches independently. When the symbolic execution is terminated,
e.g. by reaching the desired location, we use an SMT solver to derive concrete
input values from the actual path condition.

Symbolic execution works very well on programs with complicated branching
sequences \cite{Cadar08,TdH08,Pex}. Using a SMT solver symbolic execution can
effectively generate inputs systematically examining all branches of the
code. However, symbolic execution reaches its limits when confronted with a
code containing loops. Presence of even a single simple loop in a program
may cause a very large (or even infinite) number of forks in the symbolic
execution.

The problem with loops can be demonstrated on the program in
Figure~\ref{fig:runningExample}~(a). The goal is to find a feasible path to
the \texttt{assert} statement on line \texttt{9}. It is easy to see (at
least for a human) that more than one such path exists and that it must
iterate both the loops. However, there are about $2^{30}$ different execution
paths which must be explored in the worst case to show that  the code at line
\texttt{9} is reachable. The problem here is that the condition refers
 to the values of \texttt{a} and \texttt{b}, which depend on the input (the
 arrays \texttt{A} and \texttt{B}) only indirectly. Thus the condition on
 line \texttt{8} does not affect the path condition of any trace and
 therefore it is not possible for a SMT solver to compute an input leading to
 the \texttt{assert} branch. Even though this seems to be a very simple example,
it took the symbolic execution tool \Pex 99 seconds to find an
input reaching the \texttt{assert} statement. Moreover when we substituted the
predicate \texttt{a>12} on line \texttt{8} with \texttt{a>17} (thus line
\texttt{9} becomes unreachable), \Pex
was not able to finish within 5 hours.

\begin{figure*}[htbp!]
\vspace{-.5cm}
\centering
\begin{tabular}{c@{\hspace{3em}}c}
\begin{lstlisting}[language=C++]
1  int a=0, b=0;
2  for (int i=0; i<15; ++i)
3    if (A[i]==1)
4      ++a;
5  for (int j=0; j<15; ++j)
6    if (B[j]==2)
7      ++b;
8  if (a>12 && a+b==23)
9    assert(0);
\end{lstlisting}
& 
\begin{tabular}{ll}
\begin{tabular}{l}
$\chn{c_0}$ \\
\texttt{a=0} \\
\texttt{b=0} \\
\texttt{i=0} \\
\texttt{i>=15 : \{$\mathbf{c_1},\mathbf{c_2}$\}} \hspace{0.2cm} \\
\texttt{j=0} \\
\texttt{j>=15 : \{$\mathbf{c_3},\mathbf{c_4}$\}} \\
\texttt{a>12} \\
\texttt{a+b==23}
\end{tabular}
&
\begin{tabular}{ll}
\begin{tabular}{l}
$\chn{c_1}$ \\
\texttt{i<15} \\
\texttt{A[i]==1} \\  
\texttt{++a} \\
\texttt{++i}
\end{tabular}
& \hspace{0.2cm}
\begin{tabular}{l}
$\chn{c_3}$ \\
\texttt{j<15} \\
\texttt{B[j]==2} \\
\texttt{++b} \\
\texttt{++j}
\end{tabular}
\\ \\
\begin{tabular}{l}
$\chn{c_2}$ \\
\texttt{i<15} \\
\texttt{A[i]!=1} \\
\texttt{++i}
\end{tabular}
& \hspace{0.2cm}
\begin{tabular}{l}
$\chn{c_4}$ \\
\texttt{j<15} \\
\texttt{B[j]!=2} \\
\texttt{++j}
\end{tabular}
\end{tabular}
\end{tabular}
\\
(a) & (b)
\end{tabular}
\caption{Example used throughout Section~\ref{sec:Overview}. \textbf{(a)} C program
  containing loops. \textbf{(b)} Its chain program form.}
\label{fig:runningExample}
\vspace{-1cm}
\end{figure*}

\subsection{Algorithm Overview}
\label{sec:tech_overview}

We introduce our technique of handling loops on the example above, in 
three distinct phases:

\paragraph{Phase 1: Conversion to chain normal form} To better facilitate
reasoning about loops we represent the program using linear code fragments
called \emph{chains}. The decomposition of our program to chains (what we
call \emph{chain program form} later in the text) is shown in
Figure~\ref{fig:runningExample} (b). Chain $c_0$ is the topmost chain
(called \emph{root chain} later in the paper), corresponding to a path
through the code where we replace the outermost loops by constructs of the
form $\varphi$ \texttt{ : \{$c_1$, $c_2$, $\ldots$\}} with the following
meaning: at this point chains $c_1, c_2, \ldots$ may be executed any number
of times and in any order, but the condition $\phi$ must hold after we
finish executing them. Note that the condition on line \texttt{8} was
replaced by a pair of assertions.

As to the other chains, chain $c_1$ represents the path through the loop on
lines \texttt{2-4} which goes through the positive branch of the \texttt{if}
statement and $c_2$ the only other path through this loop. The same holds for
the chains $c_3$ and $c_4$ and the loop at lines \texttt{5-7}. One can easily
see that there is a natural correspondence
between the program (Figure~\ref{fig:runningExample} (a)) and its linear
representation (Figure~\ref{fig:runningExample} (b)).

The task of finding some feasible path to the \texttt{assert} statement now
depends on finding a proper interleaving of chains $c_1$ and $c_2$ for the
first loop, and $c_3$ and $c_4$ for the second one. 

\begin{figure*}[htbp!]
\vspace{-.5cm}
\centering
\begin{tabular}{c@{\hspace{3em}}c}
$\mathbf{c_1}$ & $\mathbf{c_2}$  \\
$i(\kappa_1) = \kappa_1 + \alpha_\mathtt{i}$ &
$i(\kappa_2) = \kappa_2 + \alpha_\mathtt{i}$ \\

$a(\kappa_1) = \kappa_1 + \alpha_\mathtt{a}$ &
$a(\kappa_2) = \alpha_\mathtt{a}$ \\

\multicolumn{2}{c}{$\{\mathbf{c_1},\mathbf{c_2}\}$} \\
\multicolumn{2}{c}{$i(\kappa_1, \kappa_2) = \kappa_1 + \kappa_2 + \alpha_\mathtt{i}$} \\
\multicolumn{2}{c}{$a(\kappa_1) = \kappa_1 + \alpha_\mathtt{a}$} \\
\end{tabular}
\hspace{2cm}
\begin{tabular}{cc}
$\mathbf{c_3}$ & $\mathbf{c_4}$ \\
$j(\kappa_3) = \kappa_3 + \alpha_\mathtt{j}$ &
$j(\kappa_4) = \kappa_4 + \alpha_\mathtt{j}$ \\

$b(\kappa_3) = \kappa_3 + \alpha_\mathtt{b}$ &
$b(\kappa_4) = \alpha_\mathtt{b}$ \\

\multicolumn{2}{c}{$\{\mathbf{c_3},\mathbf{c_4}\}$} \\
\multicolumn{2}{c}{$j(\kappa_3, \kappa_4) = \kappa_3 + \kappa_4 + \alpha_\mathtt{j}$} \\
\multicolumn{2}{c}{$b(\kappa_3) = \kappa_3 + \alpha_\mathtt{b}$} \\
\end{tabular}

\bigskip
\begin{tabular}{lrl}
(1) \hspace{0.5cm} & $\kappa_1 + \kappa_2     \geq 15$ \\
(2) & $\kappa_1 + \kappa_2 - 1 < 15$ & \qquad $\textrm{if}~~\kappa_1 + \kappa_2 > 0$ \\
(3) & $\kappa_3 + \kappa_4     \geq 15$ \\
\end{tabular}
\hspace{1.3cm}
\begin{tabular}{lrl}
(4) & $\kappa_3 + \kappa_4 - 1 < 15$ & \qquad $\textrm{if}~~\kappa_3 + \kappa_4 > 0$ \\
(5) & $\kappa_1                > 12$ \\
(6) & $\kappa_1 + \kappa_3     =  23$ & \qquad $\kappa_1,\kappa_2,\kappa_3,\kappa_4 \in \N$
\end{tabular}
 \caption{\textbf{Top:} Recurrent variables expressed as functions of
   counters, including the functions after merging. \newline
\hspace*{.8cm} \textbf{Bottom:} Constraint system $\mathcal{S}(c_0)$ of the root chain $\chn{c_0}$.}
\label{fig:funcsSystem}
\vspace{-.5cm}
\end{figure*}

\paragraph{Phase 2: Building a constraint system}
We start by expressing the values of variables in each chain (except root chains) as functions of the number
of times this chain was executed -- $\kappa_i$. Each chain $c_i$ is linked
to \emph{chain counter} $\kappa_i$, which takes values from $\N_0$. The link
is given by the bottom index of the counter. We show how to compute  the values
of variables on chain $c_1$, using counter $\kappa_1$.  Let
$\alpha_\mathtt{i}$ and $\alpha_\mathtt{a}$ be the initial symbolic values
of variables \texttt{i} and \texttt{a}, which are not known to this chain. Then
$i(\kappa_1) = \kappa_1 + \alpha_\mathtt{i}$ and $a(\kappa_1) = \kappa_1 +
\alpha_\mathtt{a}$ are the values of these variables expressed as functions
of $\kappa_1$. The functions for the other chains are shown in
Figure~\ref{fig:funcsSystem} (top). In terms of the original program we have
introduced a counter for each unique path through each loop.

Now for any given variable \texttt{i} and each path through a given loop
there may be different function expressing the value of \texttt{i} in terms
of the relevant counter. In the second step we try to express the value of
\texttt{i} by a single function of multiple counters.  This abstracts
from any concrete interleaving of the subchains, but the value of the
variable is expressed precisely. So in the case of chains $c_1$ and $c_2$ the value of
\texttt{i} can be expressed as $i(\kappa_1, \kappa_2) = \kappa_1 + \kappa_2
+\alpha_\mathtt{i}$.  The results for our example are presented in
Figure~\ref{fig:funcsSystem} under the headings
$\mathbf{\{c_1,c_2\}}$ and $\mathbf{\{c_3,c_4\}}$.

We can now build a constraint system for the topmost chain
$c_0$. The constraints are obtained by processing all its assertions. There  are
four assertions in the chain $c_0$: \texttt{i>=15}, \texttt{j>=15},
\texttt{a>12}, and \texttt{a+b==23}. We replace the variables by their
previously computed values (i.e. functions of counters), arriving the constraint
system  $\mathcal{S}(c_0)$ depicted at Figure~\ref{fig:funcsSystem}
(bottom).
The constraints (1), and (2) came from the assertion \texttt{i>=15}, (3),
and (4) from the assertion \texttt{j>=15}, (5) from \texttt{a>12}, and
finally (6) from \texttt{a+b==23}. The constraint (1) was computed as
follows. First we substitute variables in the assertion by their values,
obtaining $i(\kappa_1, \kappa_2) = \kappa_1 + \kappa_2 + \alpha_\mathtt{i}
\geq 15$. $\alpha_\mathtt{i}$ represents the value of \texttt{i} on reaching
the \texttt{i>=15 : \{$c_1$,$c_2$\}} instruction. Here $\alpha_\mathtt{i} = 0$,
giving us the constraint (1), which speaks about the values of $\kappa_1$
and $\kappa_2$ just after the associated loop was executed for the last
time. However, this also means that for all previous executions, where the
values are $\kappa_1'\leq\kappa_1$ and $\kappa_2'\leq\kappa_2$ such that
$\kappa_1'+\kappa_2'<\kappa_1+\kappa_2$, the negated condition $i(\kappa_1',
\kappa_2') < 15$ must hold -- i.e. there is an additional constraint for
each such choice of $\kappa_1'$ and $\kappa_2'$. This can be rephrased as
$\kappa_1+ \kappa_2 - a < 15$ for $a\in\{1,2,\ldots,\kappa_1+\kappa_2-1\}$.
Our experimentation shows that it is sufficient to take only a single 
constraint for $a=1$, giving us the constraint (2). Constraints (3), (5) and
(6) are derived similarly to (1) and constraint (4) in the same way as
(2). Note that we do not construct constraint systems for chains
$c_1, c_2, c_3, c_4$ since they do not contain any subchains.

The point of the constructed constraint system $\cS(c_0)$ is that
only those executions which reach the \texttt{assert} statement satisfy
$\cS(c_0)$ (if we instantiate the counters by the number of times
the corresponding path through a loop was executed). Which in turn means
that solving our constraint system will limit the space of counter values we
need to consider. E.g. from (5) we know that $\kappa_1\in[13,\infty]$,
and therefore the chain $c_1$, which is linked to the counter $\kappa_1$, must be executed at least
13 times on any feasible path. Similarly once we know the value for
$\kappa_1$, then  from the constraint (6) and the
fact that $\kappa_1$ is not modified in the second loop (there is no link
between $\kappa_1$ and either $c_3$ or $c_4$) we
can derive the number of times chain $c_3$ needs to be executed.

\paragraph{Phase 3: Navigating the symbolic execution}
With the chains, counters and constraint systems in place we may proceed
with the final stage of the algorithm -- finding some feasible path to line
\texttt{9}. We do this by employing slightly modified symbolic execution. We
initialize all counters to 0 and proceed down the chain $c_0$ in a standard
way until we reach the line 4: \texttt{i>=15: \{$c_1$,$c_2$\}}
(i.e. the entry point of the first loop). There are two subchains $c_1$ and $c_2$
for this loop, linked to counters $\kappa_1$ and $\kappa_2$.
Now we iteratively do the following:
\medskip

\begin{mitemize}
\item Check whether we can improve current solution of the system
  by incrementing $\kappa_1$ or $\kappa_2$. If we cannot, we stop iterating
  and continue down the chain $c_0$.
\item Otherwise we call a decision procedure to tell us which counter to
  increment. This procedure will be described in more detail in Section~\ref{sec:Algorithm}.
\item Lets assume $\kappa_1$ was chosen. In that case we symbolically
execute the chain linked to $\kappa_1$, i.e. $c_1$. We also increment the
counter $\kappa_1$.
\end{mitemize}

\noindent
Note that it may happen that the symbolic execution may get stuck because a
wrong choice of counter (counters) to increment. In
that case we backtrack, asking the decision procedure for the next best
counter to be incremented. Also note that by first establishing the values of
counters $\kappa_1$ and $\kappa_2$, before proceeding any further in the
chain $c_0$, we
significantly cut down the number of paths which need to be explored.

Having solved the loop related to chains $c_1$ and $c_2$ we proceed with the
execution, handling the loop related to chains $c_3$ and $c_4$ in the same
way. Once we arrive at the end of $c_0$ we return the current path condition,
which identifies a feasible path. In our case one such path condition is 
$\mathtt{A[0]} = 1 \wedge \ldots \wedge \mathtt{A[12]} = 1 \wedge \mathtt{A[13]} \not= 1 \wedge \mathtt{A[14]} \not= 1 ~\wedge
\mathtt{B[0]} = 2 \wedge \ldots \wedge \mathtt{B[9]} = 2 \wedge \mathtt{B[10]} \not= 2 \wedge \ldots \wedge \mathtt{B[14]} \not= 2$.
By construction this path condition is also the sought-for path condition
for the original program.

\subsection{Additional Notes}
\label{sec:OverviewNotes}

Our technique holds a significant edge over the standard symbolic execution
in proving no feasible path to a target location exists. Consider the
example in Figure~\ref{fig:runningExample} where the predicate \texttt{a>12}
on line \texttt{8} is replaced by \texttt{a>17}. 
The only change in the constraint system is (5) to (5') $\kappa_1 > 17$. 
From the constraints (2), and (5') we can derive that $\kappa_2$
has to have a negative value. Since all counters obtain values from $\N_0$, the constraint system has no solution. This means
there is no feasible path to the target location. The important fact is that
we were able to prove this even before starting the symbolic execution. 

The existence of a solution to the constraint system does not guarantee
existence of a feasible path. There are two possible problems: 1) not all
solutions correspond to feasible paths that lead to the target location, and
2) single solution may correspond to multiple paths with different
interleaving of the iterations through a loop.  So while the constraint
system effectively prunes the paths which do not lead to the target, it is
not able to give us a feasible path on its own. We therefore still need to
use the symbolic execution. For the same reasons each time we fail we need
to backtrack to check the other possible solutions.


\section{Algorithm}
\label{sec:Algorithm}

As we have explained in the previous section, the algorithm proceeds in a
three phases. 
In this section we give detailed description
of all three phases. 
Before we present our algorithm we want to state its limitations:
Currently our technique is intraprocedural -- i.e. we do not
support function calls.  We also deal only with integer variables and
arrays, and handle neither heap manipulation operations nor pointer
arithmetic.

\subsection{Phase 1: Programs as Chains}
\label{sec:AlgChains}

In this section we describe how to convert a program to chain program
form. It may be helpful for the reader to follow the example in
Fig.~\ref{fig:runningExample}.  We understand a program $P$ to be an
oriented graph, in which the vertices are the program instructions and edges
express the control flow. We write $u\to v$ ($u\to^* v$) if there is an edge
(path) from $u$ to $v$. We assume that there is a single start and a single
terminal vertex ($s_0$ and $t_0$) and that the program $P$ contains no
unreachable code. Moreover we assume that the successors of a branching
vertex are labeled \texttt{c} and \texttt{!c} (for some condition
\texttt{c}), indicating what condition must hold in order to enter the
corresponding branch.  Converting e.g. a C program to this form is obvious
(program in our definition is basically a control flow graph). We also
require the program to be in the static single assignment (SSA) form,
i.e. for each program variable there is at most one place this variable is assigned to
(using the standard conversion).

We define the \emph{chain program form C(P) of P} to be the set of all
chains in $P$.  A \emph{chain} in $P$ is a path in $P$ which is of one of
the two specific types: \emph{Root chain} is a simple path (no vertex appear
twice) $s_0\to^* t_0$. \emph{Subchain} is a simple path $v'\to^* v$ such
that it is a suffix of some path $\pi: s_0\to^*v\to v'\to^*v$ 
in $P$ where $v$ is the only vertex which appears
twice in $\pi$.  (If there are two different paths $s_0\to^* v$, then the
same path $v'\to^* v$ is treated as two different subchains.)
In the rest of the paper we treat chains as linear sequences of
vertices, and call their vertices \emph{nodes}.
In our example $\mathbf{c_0}$ is the root chain, and
$\mathbf{c_1}\ldots\mathbf{c_4}$ are the subchains.
The set of all chains can be easily obtained by unfolding the graph of $P$ 
into a tree, starting in $s_0$ and stopping each time a vertex is repeated
on a path from $s_0$ (or when we reach $t_0$).

In chains there are three types of nodes -- assume nodes, transform nodes and
loop nodes. \emph{Assume nodes}, e.g. \texttt{a>12} in $\mathbf{c_0}$, correspond to branching
conditions. \emph{Transform nodes}, e.g. \texttt{j=0} in $\mathbf{c_0}$,
correspond to assignment statements which change the programs state. Finally
\emph{loop nodes}, e.g. \texttt{i>=15 : \{$c_1$, $c_2$\}} in $\mathbf{c_0}$, are those nodes, from
which there is at least one edge in $P$ to the first vertex of some subchain. We call such a
subchain a chain \emph{associated} to this node ($\mathbf{c_1}$ and
$\mathbf{c_2}$ in this case). Note that
each subchain corresponds to a unique path through a loop and there can be
many subchains associated to the same loop node. Moreover each subchain can
also contain loop nodes, each having its own associated subchains. In
the following two phases of the algorithm we assume that there is only one
root chain. If there are multiple root chains, we run the remaining two
phases of the algorithm separately for each root chain. The results can then
be combined in an obvious way. 

Let $C(P)$ be the chain program form associated to a program $P$. Then an
\emph{execution path} in $C(P)$ is a sequence of nodes, which is created as follows:
we take some root chain and take the nodes one by one. On reaching a loop
node, we may either continue with the next node in the chain, or choose one
of the subchains associated with this loop node. In that case we take this
subchain and proceed recursively. On reaching the end in the
subchain we go ``one level up'' to the associated loop node in the parent
chain and repeat our choice to either take the next node of the parent
chain or choose another associated subchain. We finish once we reach the
terminal node for the root chain. It is easy to check that the following
statement holds:

\begin{thm} 
  The algorithm described above converts each program $P$ to chain normal
  form $C(P)$ such that for each path in $P$ there is a corresponding execution path in
  $C(P)$ and vice versa. (By correspondence we mean that the sequences of
  instructions along these two paths are the same).  Moreover if $P$ is in
  SSA form, then so is each chain of $C(P)$.
\end{thm}

\paragraph{Exponential growth of chain program form}
It can be seen that representing a program $P$ in chain program form can
bring an exponential blowup in size. Such blowup is caused by the presence
of branching statements. However, in our experience 
the number of chains is quite often very low. 
On the other hand it is not difficult to come up with a program for which the
transformation to chain program form will actually cause an exponential increase in
size. Indeed, such a growth can be observed for three of our benchmarks
Hello/HW/HWM in Section~\ref{sec:Perform}.

However, compared to vanilla symbolic execution our approach still offers significant
improvements. If we look at a symbolic execution tree of even a
very simple loop structure, we can see that every path through a loop
(represented by a single chain in our case) can appear many times in the tree -- both on
the same branch and on different branches. In other words, the size of the
chain program form is usually much smaller then the standard symbolic execution
tree for the same loop structure. Moreover, in the last stage of the
algorithm (symbolic execution of $C(P)$) the use of constraint systems
allows us to early prune branches not leading to the terminal node. So we
can have significant space and time savings over vanilla symbolic execution despite
the exponential growth of chain program form.

\subsection{Phase 2: Building the Constraint Systems}
\label{sec:AlgConstraints}

In our approach constraint systems are used to guide the symbolic execution in search for a
feasible path from the start to the target node.  Here we show how to build
the constraint system $\textsc{S}(c)$ for each chain $c$. An important idea
behind the construction is to express the values of variables used in loops
as functions of counters for the subchains. The counters in each
constraint system are linked to concrete chains. This link between
constraint systems, counters and concrete chains (in the chain program form)
is the key idea of the algorithm.

\begin{algorithm}
\dontprintsemicolon
\SetKwInOut{Input}{input}
\SetKwInOut{Output}{output}
\SetKwFunction{BuildConstraintSystem}{BuildConstraintSystem}
\Input{chain $c=i_1,i_2,\ldots, i_k$ ($i_j$ is the $j$-th instruction)}
\Output{constraint system $\mathcal{S}(c)$, and symbolic values of variables}
  $\mathcal{S}(c)$=$\emptyset$ \;
  \For{j=0; j$<$k; j++}{
    \Switch{node type of i$_j$ }{
      \Case{Loop node}{        
        \For{l=1; l$<|$subchains(n)$|$; l++}{
          \BuildConstraintSystem{l-th subchain}\;          
        }\;
        merge symbolic values returned from subchains\;
        update the symbolic state of $c$ with merged values\;
        \For{each variable \texttt{v} s.t. $c$ is the reset chain for \texttt{v}}{
           remember that $c$ is the reset chain for \texttt{v}\;
        }
      }
      \Case{Assume node}{
        instantiate the assertion $a$ (using the current symbolic state)\;
        \If{$a$ contains a counter}{
          add relevant constraints to the constraint system $\mathcal{S}(c)$ of $c$\;
        }
		 \textbf{break}\;
      }\; 
      \Case{Transform node}{
        modify the current symbolic state according to the associated
        assignment statement\;
      }
    }
  }\;
  \lFor{each variable \texttt{v} in $c$}{ 
    express the symbolic value as a function of the temporary counter $\kappa_c^\mathtt{v}$\;
  }
  \Return{symbolic values of variables}\;
  \BlankLine 
 \caption{\texttt{BuildConstraintSystem}}
  \label{fig:buildingConstraintSystem}  
\end{algorithm}

The pseudocode for this phase is shown as
Algorithm~\ref{fig:buildingConstraintSystem}. We proceed using modified
symbolic execution. The modification is twofold: First, it works on
chains, not programs. Second, the domain of symbolic values is extended to contain
counters (and expressions using counters) and a special value $\star$ with
the intended meaning ``do not know''. (Any expression containing $\star$
evaluates to $\star$.)

Let us take a chain $c$. At the beginning
each variable \texttt{i} has a symbolic value $\alpha_\mathtt{i}$ and the
constraint system $\textsc{S}(c)$ is empty. Next we 
symbolically execute the chain: Handling of the transform nodes is
clear. Assume nodes are treated as sources of constraints for $\textsc{S}(c)$.  Each assertion
is first instantiated with the current  values of variables, and
then inserted to the constraint system only if it references some
counter. (As per the example in Section~\ref{sec:tech_overview} multiple
constraints can be produced from a single assertion.)  On reaching a loop
node $n$ we first recursively build the constraint systems for all subchains associated
to this node, obtaining symbolic values of variables (which can now depend
on counters of some (possibly nested) subchains). For each
variable we then merge the symbolic values obtained in the
subchains (see the section \emph{Merging values ...} below). The current
symbolic state of $c$ is then updated with the merged values.
At this point we also detect the variables for which this chain is the reset
chain (see the section \emph{Expressing values ...} below).
Since each loop node has an associated branching condition, we finish
processing this condition as we would for the assume node. Finally, when we
reach the end of the chain, we express the values of variables as
functions of loop counters (and return these values). $\textsc{S}(c)$ now contains the complete constraint
system for the chain $c$.
We now describe the process in more detail. We start by explaining the last
step, because it is here where counters are dealt with and the notion of
recurrent variables introduced.

\paragraph{Expressing values using counters}
Let us fix a chain $c$. The goal is for each variable to compute a function expressing
its value in terms of counters. 
We focus on so called \emph{recurrent variables}, which are the variables
whose value 1) changes on the execution path corresponding to $c$ and 2)
their value is function of their initial value before executing $c$. An
example of a recurrent variable is the variable $\mathtt{i}$ in the chain
$c_1$, for which we get $\mathtt{i}=\alpha_\mathtt{i}+1$. To detect
recurrent variables for a given chain $c$ we simply analyze symbolic state
resulting from symbolic execution of this chain.

For each recurrent variable we express its value in terms of how many
times the chain $c$ was executed -- using the counter $\kappa_c$ associated
with the chain $c$. In our example,
$i(\kappa_1)=\alpha_\mathtt{i}+\kappa_1$.  We use a very simple custom difference
equation solver to express the values of variables using counters. 
Our solver handles only those recurrences which
correspond to arithmetic (e.g. $\mathtt{i}=\alpha_\mathtt{i}+7$) and geometric
(e.g. $\mathtt{i}=3\cdot\alpha_\mathtt{i}$) progressions. This restriction is
justified by the fact that, according to our experience, overwhelming
majority of code uses only such progressions. In case we are not able to
solve a recurrence, we use the ``do not know'' value $\star$. 
 
An important point to make is that the initial value for \texttt{i},
$\alpha_\mathtt{i}$, can be set by some chain $r$, of which the current
chain $c$ is a subchain. Therefore the value of \texttt{i} does not depend
only on the number of times $c$ was executed, but, more specifically, on the number of times $c$
was executed since last execution of $r$. Therefore the value of \texttt{i}
in fact depends on a counter $\kappa_c^{r}$ parametrized by two chains:
the \emph{update chain} $c$ and the \emph{reset chain} $r$ --
i.e. $i(\kappa_c^{r})=\alpha_\mathtt{i}+2\cdot\kappa_c^{r}$. This counter
is incremented each time the chain $c$ is executed, and set to zero each
time the chain $r$ executed. If there is no reset chain for a given
variable, we use the plain counter $\kappa_c$, where $c$ is the update
chain and the root chain is used as the reset chain. Note that all counters used in our example in
Figure~\ref{fig:runningExample} are of this type. The following statement is
true for chain program forms, and follows from the fact all chains are in
the SSA form:

\begin{lem}
  Let \texttt{v} be a recurrent variable whose update chain is $c$. Then
  \texttt{v} is not reset (to its initial value) in any subchain of $c$ and
  there exists at most one superchain of $c$ where \texttt{v} is reset.
\end{lem}

At the time of processing the update chain $c$ for \texttt{v} we do not know
yet what superchain of $c$ is the reset chain for \texttt{v}. Therefore we
use a temporary counter $\kappa_c^\mathtt{v}$ to express the value of
\texttt{v}. When processing a chain $d$ such that 1) $d$ is a superchain of
$c$ and 2) the value of \texttt{v} is no longer given by a recurrence
expression ($\alpha_\mathtt{v}$ does not occur in the symbolic value of
\texttt{v}), we know that $d$ is the reset chain for \texttt{v}. We remember
this information, and once all constraint systems are built we replace all
occurrences of each temporary counter $\kappa_c^\mathtt{v}$ in all constraint
systems with the correct counter $\kappa_c^d$.

\paragraph{Merging values from subchains}
We explain the merging process on the case of two subchains. The extension
to multiple subchains is straightforward.

Let us assume that a chain has two subchains with the associated counters
being $\kappa_c$ and $\kappa_d$ (as the reset chains are not important
here, we omit the upper indices) and there is a variable \texttt{i} value of which
is expressed as $i=i_1(\kappa_c)$ in the first chain and $i=i_2(\kappa_d)$
in the second. We would like to ``merge'' the values of \texttt{i} -- i.e. to find
a function $i(\cdot,\cdot)$ such that $i=i(\kappa_c,\kappa_d)$. Let $\alpha_\mathtt{i}$
be the symbolic value of \texttt{i} on entering the subchains. There are
some simple cases: e.g. if $i_1(\kappa_c)=i_2(\kappa_d)=v$ for some constant
$v$, then obviously also $i(\kappa_c,\kappa_d)=v$. Similarly if
$i_1(\kappa_c)=i_2(\kappa_d)=\alpha_\mathtt{i}$. On the other hand if
$i_1(\kappa_c)=v_1\not=v_2=i_2(\kappa_d)$ then there is no such function
$i(\kappa_c,\kappa_d)$. In that case we put $i(\kappa_c,\kappa_d)=\star$.

The most interesting case is when both $i_1(\kappa_c)$ and $i_2(\kappa_d)$
depend on $\alpha_\mathtt{i}$ -- e.g.  $i_1(\kappa_c) = v_1\cdot\kappa_c +
\alpha_\mathtt{i}$ and $i_2(\kappa_d) = v_2\cdot\kappa_d + \alpha_\mathtt{i}$. This means that
the value of \texttt{i} is updated in both subchains. In this case we can easily
derive that $i(\kappa_c,\kappa_d)=v_1\cdot\kappa_c + v_2\cdot\kappa_d +
\alpha_\mathtt{i}$. Table~\ref{tab:merging} sums up all supported merge operations
for functions of a single counter. In all other cases we put
$i(\kappa_c,\kappa_d)=\star$.  
From Table~\ref{tab:merging} we see that our ability to merge is
limited. E.g. we are not able to merge even $i_1(\kappa_c) =
v_1\cdot\kappa_c + \alpha_\mathtt{i}$ and $i_2(\kappa_d) =
\alpha_\mathtt{i}\cdot v_2^{\kappa_d}$. Also $\star$ propagates quickly
through the system and constraints with $\star$ are not useful in our
approach.
On the other hand we can merge functions with
different numbers of counters, e.g.  $i(\kappa_c,\kappa_d)$ with
$i(\kappa_3)$ etc.

\begin{table}[!htbp]
  \centering
  \begin{tabular}{|c|c|c|}
    \hline
    $i_1(\kappa_c)$ & $i_2(\kappa_d)$ & $i(\kappa_c,\kappa_d)$ \\
    \hline
    $\alpha_\mathtt{i}$ & $\alpha_\mathtt{i}$ & $\alpha_\mathtt{i}$ \\
    $v$ & $v$ & $v$\\
    $v_1\cdot\kappa_c + \alpha_\mathtt{i}$ & $v_2\cdot\kappa_d + \alpha_\mathtt{i}$ & 
    $v_1\cdot\kappa_c + v_2\cdot\kappa_d + \alpha_\mathtt{i}$  \\
    $\alpha_\mathtt{i}\cdot v_1^{\kappa_c}$ & $\alpha_\mathtt{i}\cdot v_2^{\kappa_d}$ & 
    $\alpha_\mathtt{i}\cdot v_1^{\kappa_c}\cdot v_2^{\kappa_d}$ \\
    \hline 
  \end{tabular}
\medskip

  \caption{Supported merge operations for unary functions}
  \label{tab:merging}
\vspace{-1.2cm}
\end{table}

\subsection{Phase 3: Constraints-Driven Symbolic Execution}
\label{sec:AlgSolving}

The last stage of our algorithm is to navigate (modified) symbolic execution
in order to find a feasible path from $s_0$ to $t_0$. We
modify standard symbolic execution in order to run on the chain program form
described in Section~\ref{sec:AlgChains}. To do so, we first extend the
symbolic state by extra variables representing the values of
counters. Second, on entering a chain, we instantiate all symbols $\alpha_\mathtt{v}$ in the
constraint system associated with the chain by their
actual symbolic values.

The symbolic execution starts by setting all counters for which the current
chain is the reset chain to zero and then proceeds on the root chain as
normal until it reaches a loop node, which will play a role of a branching
statement in standard symbolic execution.  A symbolic execution tool
typically asks an oracle (a heuristic), when an execution reaches forking
branch. Since two or more branches can be simultaneously taken from that
point, an oracle is responsible to choose a branch which is more likely to
reach the goal of exploration then others. In our case branching points are
the loop nodes, and the oracle is the decision
procedure given by Algorithm~\ref{alg:chooseChain}.

\begin{algorithm}
\dontprintsemicolon
\SetKwInOut{Input}{input}
\SetKwInOut{Output}{output}
\SetKwFunction{chooseChain}{chooseChain}
\Input{$c,D$~:: a chain and an subset of its sub-chains\\$A$~:: constraint system for $c$}
\Output{Chosen chain (i.e. $c$ or some $d\in D$) or {\bf null}.}
\lIf{$A$ has no solution}{
  \Return{\bf null}\;
}
\lIf{counters' values represent a solution of $A$}{
  \Return{c}\;
}
$R$ := $\{$reset chains of counters for (subchains of) $D$, whose reset gets closer to a solution of $A\}$\;
$U$ := $\{$update chains of counters for (subchains of) $D$, whose update gets closer to a solution of $A\}$\;
\lIf{$R\cup U=\emptyset$}{
  \Return{c}\;
}\lElse{
\Return{arbitrary element from $R\cup U$}
}\;
\BlankLine
\caption{\texttt{chooseChain}}
\label{alg:chooseChain}
\end{algorithm}

Let $c$ be the currently executed chain, $A$ its (instantiated) constraint
system, $i$ the processed loop node, and $D$ be the
subset of the set of subchains associated to $i$ (containing those subchains
which have not been yet explored during backtracking). If $A$ has no
solution, we immediately stop symbolic execution for this branch. Otherwise,
if the current values of counters already form a solution of $A$, we continue executing
$c$, as there is no reason to execute any of the subchains. Otherwise we need to choose a chain
$d\in D$ which, hopefully, brings us closer to a solution of $A$. If there
is such $d$, we continue with the symbolic execution of $d$. Finally if there
is no such $d$, then we also continue executing $c$, hoping that we can closer to
a solution of $A$ at some loop node below.

Now we describe what we mean by ``getting closer to a solution of A''. Let
$\vec{w}$ be a vector of current values of all the counters such that
$\vec{w}$ is not a solution to $A$. We now ask whether there is a vector
$\vec{v}$ on natural numbers such that 1) $\vec{v}+\vec{w}$ is a solution to
$A$, and 2) there is a counter $\kappa$ such that $d\in D$ (or some of its
subchains) is the update chain for $\kappa$ (reset chain for $\kappa$) and
there is a positive (negative) number in in the corresponding position in
$\vec{v}$. If yes, then executing the chain $d$ gets us ``closer to a
solution of $A$''.

There are many possible approaches to compute the vector $\vec{v}$. One is
to simply use a SMT solver to obtain a solution to $A$. In our
implementation we use interval abstraction: we overapproximate the set of
all solutions by giving a
set of intervals for each counter. These intervals are derived from
the constraint system, and we have a solution to $A$ if the value of each counter
lies in one of its intervals. In this abstraction individual components of
any solution vector are independent. Thus choosing some vector $\vec{v}$ is
trivial.

Finally we have to say what happens when the symbolic execution reaches the
terminal node of a chain $c$. We first increment all the associated
counters $\kappa_c^d$ (for all $d$). If $c$ is a subchain we continue by
(again) executing the associated loop node in the parent chain, otherwise $c$ is a root
chain and we reached the target node.

We conclude by stating the soundness and incompleteness of our method (the
latter follows immediately from incompleteness of the standard symbolic execution):

\begin{thm}[Soundness]
  If the symbolic execution of $C(P)$ (as described in
  Section~\ref{sec:Algorithm}) terminates with success,
  then the returned path  condition represents a feasible path from start to target
  instruction in the original program $P$. Moreover if the symbolic
  execution fails, then there is no feasible path in $P$ to the target instruction.
\end{thm}

\begin{thm}[Incompleteness]
  There exists a program $P$ with
reachable target instruction for which the symbolic execution of $C(P)$ never terminates.
\end{thm}


\section{Experimental Results}
\label{sec:Results}

To evaluate the effectiveness of our technique we implemented it (with all
the restrictions mentioned at the beginning of Section~\ref{sec:Algorithm}) in our tool
\CBATool, and tested it on a set of nine benchmarks. We  also compared the
performance of \CBATool to that of two very successful tools \Pex\cite{TdH08,Pex}
and \Klee\cite{CDE08}. All the nine benchmarks share some common
properties: 1. the code contains loops (so the benchmarks produce a huge symbolic execution
tree) 2. there is a unique location to be reached
3. they consist of only one function (since
our technique does not handle function calls). In the  first six
benchmarks the goal is to find a feasible path to the target location.
On the other hand in the last three benchmarks there is no
feasible path to the target location and the goal is to show that no
feasible path exists.

\paragraph{Benchmark Description}
The first three benchmarks \textbf{Hello/HW/HWM} are adapted from \cite{AGT08}~(there is only
verbal description, no code). The HWM benchmark accepts a \texttt{C} string
as an input and scans the string for the presence of substrings
\texttt{"Hello"},\texttt{"World"}, \texttt{"At"} and \texttt{"Microsoft!"}.
HW and Hello are simplified versions of the HWM benchmark, looking for the
first two words (one word) only. 

In \textbf{DOIF} we model a typical piece of code which scans an input and,
for each member of the input array, performs an action which depends on its
value. This benchmark is supposed to exercise primarily the third stage of
the algorithm.  Branching inside the loops enormously expands the number of
paths in the model. \textbf{DOIFex} is an extension of this benchmark, and
tests behaviour on sequences of loops with internal branching.

The \textbf{EQCNT} benchmark contains nested loops with branching, where a
variable defined in the outermost scope is modified in the innermost
loop. \textbf{EQCNTex} is a modified benchmark (in a sense two instances of
EQCNT in sequence), however the number loop iterations is now given
explicitly (in contrast to the two remaining benchmarks, where it is
dependent on the input). For an algorithm to be efficient on this benchmark
it has to aggressively prune infeasible paths.

The \textbf{OneLoop} benchmark consists of simple loop in which the variable
\texttt{i}, with initial value $0$, is increased by $4$ in every
iteration. Once the loop is finished we check whether \texttt{i==15}, which
is false for any value of the input variable \texttt{n}. 
 \textbf{TwoLoops} is a an
extension of the previous benchmark by adding a second loop, whose
loop condition depends on the value computed in the first loop.

\subsection{Tool Comparison}
In this section we present the experimental results we obtained by running
\Pex, \Klee and our tool \CBATool on our set of benchmarks.
 We ran our test on an Intel i7/920 2.67GHz Windows machine with 6GB
of RAM. 
Since \Klee is native C++ Linux application, we used the Cygwin library to
run \Klee on Windows, resulting in an overhead caused by calls to Cygwin's dynamic library. We
decided to reduce this negative effect by using the \texttt{time} utility to
measure the 'user' time of \Klee. However this was as close as we could get
to running all the tools in the same environment.

For \Pex we present two results for each of the benchmarks. This is because
the performance of \Pex is affected by a set of configurable
parameters. The first result is obtained in the way recommended by \Pex
developers -- with all parameters set to infinity. The second result
(indicated by an asterisk), which
is usually better, is obtained by iteratively running \Pex and adjusting the
parameters according to suggestions provided with the unsuccessful runs.

\paragraph{Comparison results}
The results are presented in Table~\ref{TabExprRes}. We measured the time
required to reach the target location. Each benchmark has an associated
timeout (column \tout), which was set according to the perceived difficulty of that
particular benchmark. The success was defined as reaching the target
location (or demonstrating it is not possible to reach this location) within
the specified time limit.

Looking at the table one can see that, on our set of benchmarks, \CBATool
significantly outperforms both \Pex and \Klee. This shows that on short
pieces of code containing non-trivial loops our technique can effectively
guide the symbolic execution to the chosen target location. On the other
hand \Pex and \Klee clearly suffer from the limitations of the symbolic
execution when dealing with loops.

\begin{table}[!htb]
\vspace{-.3cm}
\centering
    \begin{tabular}{||*{6}{c|}|}
      \hline

      \bf Test  &
      \tout &
      \Pex &
      \Pex$^*$ &
      \Klee &
      \CBATool \\
      \hline
      \hline
      Hello    & 30m & 3.234s  & 7.233s  & 0.093s           & 0.026s \\
      HW       & 1h  & 14.890s & 11.107s & 37m 0s           & 0.175s \\
      HWM      & 1h  & \textbf{fail}     & 8m 54s           & \tout & 1.997s \\
      DOIF      & 30m & \tout  & 20m 27s          & \tout & 0.388s \\
      DOIFex       & 1h  & \tout  & \tout & \tout & 1.745s \\
      EQCNT    & 30m & 1m 43s  & 11.592s & \tout & 0.191s \\
      \hline
      EQCNTex  & 1h  & 46m 12s & 42m 20s & \tout & 2.458s \\
      OneLoop  & 30m & 2m 14s  & 4m 27s  & \tout & 0.002s \\
      TwoLoops & 30m & 1m 4s   & 57.426s & \tout & 0.003s \\
      \hline
    \end{tabular}

\medskip
  \caption{Running times of \Pex, \Klee and \CBATool.}
\label{TabExprRes}
\vspace{-1.2cm}
\end{table}

\subsection{Performance Analysis of \CBATool}
\label{sec:Perform}

In this section we discuss the behaviour of \CBATool on our set of nine
benchmarks. Table~\ref{TabCBProfile} shows the performance data. The three
enclosing columns refer to the three stages of our algorithm: \emph{Chain
  prog. form} refers to the conversion of a program into chain program
form. \emph{Chains} gives the number of root/all chains, \emph{Time} the
time\ needed for the conversion and \emph{Space} the size of resulting data
structures. \emph{Constr. Systems} covers the second stage.
\emph{Elim} shows how many root chains were shown to be
infeasible even before getting to the last stage and \emph{Size} gives the
total number of constraints left after pruning. Finally
\emph{Constraints-Driven Sym. Exe.} refers to the last stage.
\emph{SStat} gives the number of symbolic states (i.e. vertices of a
symbolic execution tree) visited. \emph{CSOL} gives the number of calls to
the constraint solver: The first number is for the initial solution, the
second one for the remaining calls. \emph{SMT} is the number of calls to the
Z3 SMT solver and finally \emph{PC} counts the number of predicates in the
resulting path condition. 

\begin{table*}[!htb]
\centering
\vspace{-.3cm}
\begin{tabular}{||*{13}{c|}|}
\hline
&
\multicolumn{3}{c|}{\bf{Chain Prog. Form}} &
\multicolumn{4}{c|}{\bf{Constr. Systems}} &
\multicolumn{5}{c||}{\bf{Constraints-Driven Sym. Exe.}} \\
\cline{2-13}
\bf Test &
Chains & Time & Space &
Elim & Size & Time & Space &
SStat & CSOL & SMT & Time & PC \\
\hline
\hline
Hello & 3/6 & 0.003s & 1kB & 2 & 5 & 0.001s & 3.1kB & 10 & 8 / 36 & 19 & 0.023s & 5 \\
HW & 9/17 & 0.009s & 20 kB & 8 & 10 & 0.040s & 6 kB & 44 & 30 / 170 & 92 & 0.144s & 10 \\
HWM & 81/161 & 0.097s & 369 kB & 80 & 20 & 0.719s & 12 kB & 174 & 112 / 686 & 376 & 1.456s & 22 \\
DOIF & 1/5 & 0.003s & 1 kB & 0 & 3 & 0.014s & 2 kB & 98 & 97 / 349 & 136 & 0.380s & 26 \\
DOIFex & 1/9 & 0.004s & 3 kB & 0 & 4 & 0.008s & 5 kB & 211 & 209 / 728 & 212 & 1.757s & 26 \\
EQCNT & 1/4 & 0.004s & 1 kB & 0 & 3 & 0.003s & 2 kB & 45 & 44 / 245 & 45 & 0.187s & 43 \\
\hline
EQCNTex & 1/7 & 0.004s & 2 kB & 0 & 6 & 0.005s & 4 kB & 1192 & 1022 / 7286 & 1233 & 2.458s & 0 \\
OneLoop & 1/2 & 0.003s & 293 B & 0 & 2 & 0.001s & 698 B & 0 & 1 / 0 & 0 & 0.001s & 0 \\
TwoLoops & 1/3 & 0.003s & 578 B & 0 & 2 & 0.001s & 1 kB & 0 & 1 / 0 & 0 & 0.001s & 0 \\
\hline
\end{tabular}

\medskip
\caption{Performance data for \CBATool on our set of benchmarks.}
\label{TabCBProfile}
\vspace{-1cm}
\end{table*}

The number of chains for the HWM benchmark is quite large.
 There were 161 chains, including 81 root chains.
The high number of chains for HWM is reflected in the time and space needed
to build the chain program form. We can see the negative effect of exponential
growth of chain program form here. If we compare the running time and the number
of chains for the three related benchmarks Hello/HW/HWM we see that number of chains
grow indeed exponentially.

Another interesting observation is that it is hard to predict how long will
the last stage take based on the performance of the first two
stages. To see this, consider the results obtained for the HWM and DOIFex
benchmarks. In the case of HWM there are $161$ chains (before pruning) and 20
constraints in remaining chains (after pruning), while for DOIFex there are only 9 chains and 4
constraints. However in both cases the last stage explores a comparable
number of symbolic states in similarly comparable time. This indicates that,
in respect to our algorithm, the number of chains and constraints are not
the only important parameters.

The last negative output can be seen on the EQCNTex benchmark. The
values for the last stage are an order of a magnitude higher than for
the other tests. Note that this is despite the number of chains being very
low. Remember that in EQCNTex there is no feasible path to the target
location, and many paths need to be explored to prove it. EQCNTex
benchmark shows the limitations of our algorithm with respect to solving such
problems. Even though it can effectively prune away many paths (as witnessed
by the last two benchmarks, OneLoop and TwoLoops) this is not always
sufficient. Nevertheless \CBATool still fared significantly better on EQCNTex than both
\Pex and \Klee.


\section{Related Work}
\label{sec:Related}
The earliest work dealing with symbolic execution~\cite{BEL75,Kin76}
showed that symbolic execution can be an effective approach to test
generation. However the astronomical blowup of program model caused by loops
was not in the centre of interest. Usability evaluation of
symbolic execution for proving correctness of program implementing Floyd's
method~\cite{Floyd67} was in~\cite{Kin76}, but
problems with loops were handled by manually inserting \texttt{ASSUME}
statements where necessary.

Modern effective techniques based on symbolic execution are mostly hybrid, combine
symbolic execution with some other approaches. The first group are techniques based of
combining (alternating) concrete and symbolic
execution~\cite{PKS05,SMA05,TdH08,GLM08:fuzzing}. This approach primarily
avoids the problems caused by limitations of SMT solvers. Although the
practical usability is greatly improved, these techniques have no effect on
the ability to handle loops. The second group combines symbolic execution
with some validation
technique~\cite{GNRT10,GMR09,Beckmanetal08,NRTT09,Gulavanietal06}. This
approach is much more successful from the point of handling loops. Thanks to
employing the complementary techniques, many symbolic paths can be effectively
pruned away when exploring the symbolic state space. This can often lead to
effective navigation of symbolic execution in programs with loops.
There is also a group of techniques which aim to make symbolic execution
effective in the general case, not specifically focused on just programs
with loops~~\cite{BCE08,G07,AGT08,Cadar08,CDE08,GLM08:active_props}.

The idea of using constraint system for analyzing loops was considered
before in different contexts. First approach, dating back to 70's, infers
relations between program variables~\cite{K76:affine_vars,CH78}, while the
more recent techniques are primarily focused on formal verification, and
inductive invariant computation~\cite{BHMR07,GSV08}. 
 Analysis of loops using loop-counters as the
artificial program variables is also well known~\cite{progAnalysis}.

The technique of Loop-Extended Symbolic Execution~\cite{SPmCS09} (LESE) is
probably the one most closely related to our approach. The LESE approach
introduces symbolic variables for the number of times each loop was
executed, and links these with features of a known input grammar such as
variable-length or repeating fields. This allows the symbolic constraints to
cover a class of paths that includes different number of loop iterations,
expressing loop-dependent program values in terms of properties of the input.

Our approach is very different: Instead of extending the input by new
symbolic variables to reason about multiple symbolic execution paths at
once, our goal is to build a constraint systems to steer the symbolic execution through loops towards a
specified target. For this reason we introduce counters which are linked to
different paths through a cycle, contrasting to the overall iteration count
used by the LESE approach. Our technique therefore applies to a much more
general class of programs.

Finally there is an orthogonal line of research which tries to improve the
symbolic execution for programs with some special types of inputs. Some
examples are techniques for dealing with programs with
string inputs~\cite{BTV09,XGM08}, and techniques which reduce input space
given by an input grammar~\cite{GKL08,SPmCS09}. These approaches can be
effective on loops when such loops are closely related to the input.


\section{Conclusion and Future Work}
\label{sec:Conclusions}

In this paper we introduced a new algorithm for effective navigation of symbolic
execution through loop containing code. The algorithm infers a collection of
constraint systems and uses them to steer the symbolic execution towards a target
location. To build these constraint systems we express the values of
variables modified in a loop as functions of the number of times a
particular path through the  loop was executed. 

We have also built an experimental implementation of our technique and tested its
effectiveness on a set of nine benchmarks. Our tool was able to correctly
solve each of these benchmarks within seconds, being several orders of
magnitude faster than the leading symbolic execution tools. Moreover we have
demonstrated that our technique is also useful for proving that no feasible
path to a target location exists.

Finally, we argue that it would be beneficial for general-purpose tools
based on symbolic execution to integrate our technique as a new search
strategy. This strategy would then be activated each time the symbolic
execution needs to navigate to  a specific target location below some
complicated loop structure. Since our algorithm is itself based on symbolic
execution, such integration should not be too difficult.

There are many interesting open directions for future work. An obvious task
would be to extend our approach to interprocedural setting.  Moreover, so far
we have considered only integer variables and arrays. It would be
interesting to extend our technique to handle more
sequential containers (e.g. lists or vectors) and/or floating point arithmetics. Another approach is to try to
curb the growth of the chain program form, for example by merging those
chains which have the same effect on program execution. Finally it would be
nice to actually integrate our approach with the existing symbolic execution
tools like \Klee.



\paragraph{Acknowledgements}
We would like to thank Nikolai Tillmann for promptly answering our questions
regarding \Pex. We also thank V\'aclav Bro\v{z}ek, Vojt\v{e}ch Forejt,
Anton\'\i n Ku\v{c}era  and
Jan Strej\v{c}ek for their helpful comments on earlier drafts of this paper.


\bibliographystyle{plain}
\bibliography{verification}


\end{document}